\documentclass{appolb}
\usepackage{epsfig}

\begin{document}

\title{The perturbation method to solve subdiffusion--reaction equations%
\thanks{Presented at 24th Marian Smoluchowski Symposium on Statistical Physics}%
}

\author{Katarzyna D. Lewandowska
\address{Department of Physics and Biophysics, Medical University of
         Gda\'nsk,\\ ul. D\c{e}binki 1, 80-211 Gda\'nsk, Poland.}
\and
Tadeusz Koszto{\l}owicz, Mateusz Piwnik
\address{Institute of Physics, Jan Kochanowski University,\\
         ul. \'Swi\c{e}tokrzyska 15, 25-406 Kielce, Poland.}
}

\maketitle

\begin{abstract}
We use the perturbation method to approximately solve subdiffusion--reaction equations. Within this method we obtain the solutions of the zeroth and the first order. The comparison our analytical solutions with the numerical results shown that the preturbation method can be useful to find approximate solutions of nonlinear subdiffusion--reaction equations.
\end{abstract}


\section{Introduction}

There are a lot of differential equations which have solutions only in a very few special cases and their general solutions remain unknown. Such an example is nonlinear differential equations with a fractional time derivative which describes the subdiffusion--reaction symmetrical system with two initially separated diffusing particles of speices ${\rm A}$ and ${\rm B}$ reacting according to the formula ${\rm A}+{\rm B}\rightarrow \emptyset(inert)$ \cite{ya,yal,kl1}
\begin{eqnarray}
  \frac{\partial^{\alpha}A(x,t)}{\partial t^{\alpha}}&=&D\frac{\partial^2 A(x,t)}{\partial t^2}-k A(x,t)B(x,t)\;,\label{eq3:a}\\
  \frac{\partial^{\alpha} B(x,t)}{\partial t^{\alpha}}&=&D\frac{\partial^2 B(x,t)}{\partial t^2}-k A(x,t)B(x,t)\;,\label{eq3:b}
\end{eqnarray}
where $A$ and $B$ denote the concentrations of diffusing particles of species ${\rm A}$ and ${\rm B}$, respectively, $D$ --- their subdiffusion coefficients (the same for both substances), $k$ is the reaction rate constant and the Caputo fractional derivative $\partial^\alpha f(x)/\partial t^\alpha$  is defined as \cite{p}
\begin{equation}
  \label{c}
  \frac{\partial^{\alpha}f(t)}{\partial t^{\alpha}}=\frac{1}{\Gamma(n-\alpha)}\int_{0}^{t}dt' \frac{f^{(n)}(t')}{(t-t')^{1+\alpha-n}}\;,
\end{equation}
$f^{(n)}$ denotes the derivative of natural order $n$ and $n-1\leq\alpha<n$.
We assume that the substances are separated from ech other at an initial moment and
the initial conditions are
\begin{equation}
  \label{wp}
  A(x,0)=\left\{ \begin{array}{ll}
      C_{0}\;, & x<0\;, \\
      0\;, & x>0\;,
    \end{array} \right.\qquad
  B(x,0)=\left\{ \begin{array}{ll}
      0\;, & x<0\;, \\
      C_{0}\;, & x>0\;.
    \end{array} \right. 
\end{equation}          
The symmetry of the system gives $A(-x,t)\equiv B(x,t)$.

Since the general method of solving fractional subdiffusion--reaction equations has not been found yet, one usually uses various approximations, such as the quasistationary approximation \cite{kl1} or the scaling method \cite{yal}. In this paper we present an idea to solve differential equations by means of the perturbation method. The perturbation method is usually used in nonlinear equations in the following form
\begin{equation}
  \label{eq:1}
  \frac{\partial f}{\partial x}=\mathcal{F}(f)+\epsilon\mathcal{G}(f)\;,
\end{equation}
where $\epsilon$ is a dimensionless small parameter and it is assumed that the solution of the equation 
\begin{equation}
  \label{eq:1a}
  \frac{\partial f_0}{\partial x}=\mathcal{F}(f_0)\;,
\end{equation}
is known. Then, the solution of Eq.~(\ref{eq:1}) is a power series with respect to the parameter $\epsilon$
\begin{equation}
  \label{eq:2}
  f=\sum_{n=0}^{\infty}\epsilon^n f_n\;.
\end{equation}
However, the perturbation method cannot be used directly in Eqs.~(\ref{eq3:a}) and~(\ref{eq3:b}). The reason is that the system described by Eqs.~(\ref{eq3:a}), (\ref{eq3:b}) and~(\ref{wp}) has its own structure where the depletion zone, the reaction region and diffusion region occur. As is shown in \cite{kl1} within the reaction region the terms on the right hand side of Eq.~(\ref{eq3:a}) (or~(\ref{eq3:b})) are comparable to each other and do not fulfil the assumptions of the perturbation method. Let us note that the reaction term $kAB$ and the diffusion term $D\partial^2 \{A,B\}/\partial x^2$ have different scaling properties \cite{yal}. We use this scaling property and we change the variables ($x$ and $t$) to the dimensionless variables in such a way that the reaction term will be small compared to the diffusion term. Then, we need to solve the transformed equations by means of the perturbation method, and next we return to the original variables. We expect that the perturbation method works for the transformed dimensionless equations for $\epsilon\ll 1$ but it is not obvious that the dimensional approximate solutions are accurate. To check the correctness of these solutions we will compare them with numerical solutions to Eqs.~(\ref{eq3:a}) and~(\ref{eq3:b}). In our paper we will find the solution of the zeroth and the first order of Eqs.~(\ref{eq3:a}) and~(\ref{eq3:b}). As far as we know the perturbation method has not been yet applied to solve fractional subdiffusion--reaction equations. We add that the preturbation method was appiled to solve normal diffusion--reaction equations with one static and one mobile reactants \cite{lk}.

\section{The perturbation method}

Firstly, we transform the equations (\ref{eq3:a}) and (\ref{eq3:b}) to the dimensionless form using the substitutions
\begin{equation}
  \label{}
x=\rho x_s\;,\quad t=\tau t_s\;,
\end{equation}
where $\rho$ and $\tau$ denote the dimensionless position and time, $x_{\rm s}$ and $t_{\rm s}$ are constants of the dimension of space and time, respectively. We obtain
\begin{eqnarray}
\frac{\partial^\alpha a(\rho,\tau)}{\partial \tau^{\alpha}}&=&\frac{\partial^2 a(\rho,\tau)}{\partial \rho^{2}}-\epsilon a(\rho,\tau)b(\rho,\tau)\;,\label{eq:2a}\\
\frac{\partial^\alpha b(\rho,\tau)}{\partial \tau^{\alpha}}&=&\frac{\partial^2 b(\rho,\tau)}{\partial \rho^{2}}-\epsilon a(\rho,\tau)b(\rho,\tau)\;,\label{eq:2b}
\end{eqnarray}
where 
\begin{displaymath}
  a(\rho,\tau)=\frac{A(\rho x_s,\tau t_s)}{C_{0}}\;,\quad b(\rho,\tau)=\frac{B(\rho x_s,\tau t_s)}{C_0}\;,
\end{displaymath}
\begin{equation}
  \label{eq:4}
  \quad x_s=\sqrt{Dt^{\alpha}_s}\;,\quad \epsilon=kt^{\alpha}_s C_0\;,
\end{equation}
with the initial conditions
\begin{equation}
  \label{wp_a}
  a(\rho,0)=\left\{ \begin{array}{ll}
      1\;, & \rho<0\;, \\
      0\;, & \rho>0\;,
    \end{array} \right.\qquad
  b(\rho,0)=\left\{ \begin{array}{ll}
      0\;, & \rho<0\;, \\
      1\;, & \rho>0\;.
    \end{array} \right. 
\end{equation}          
and the boundary ones
\begin{equation}
  \label{wb_a}
  a(-\infty,\tau)=b(\infty,\tau)=1\;, \qquad \left.\frac{\partial a(\rho,\tau)}{\partial \rho}\right|_{\rho=0}=\left.\frac{\partial b(\rho,\tau)}{\partial \rho}\right|_{\rho=0}\;.
\end{equation}

We assume that the solutions to Eqs.~(\ref{eq:2a}) and~(\ref{eq:2b}) are given in the form of~(\ref{eq:2})
\begin{equation}
  \label{eq:12}
  a(\rho,\tau)=\sum_{n=0}^{\infty}\epsilon^{n}a_{n}(\rho,\tau)\;,\qquad b(\rho,\tau)=\sum_{n=0}^{\infty}\epsilon^{n}b_{n}(\rho,\tau)\;.
\end{equation}

Substituting Eqs.~(\ref{eq:12}) to Eqs.~(\ref{eq:2a}) and (\ref{eq:2b}) and comparing the functions of the same order with respect to the variable $\epsilon$ occurring on the both sides of these equations we obtain the following equations for the functions of the zeroth order
\begin{eqnarray}
  \frac{\partial^\alpha a_0(\rho,\tau)}{\partial \tau^{\alpha}}&=&\frac{\partial^2 a_0(\rho,\tau)}{\partial \rho^{2}}\;,\label{eq:5a}\\
  \frac{\partial^\alpha b_0(\rho,\tau)}{\partial \tau^{\alpha}}&=&\frac{\partial^2 b_0(\rho,\tau)}{\partial \rho^{2}}\;,\label{eq:5b}
\end{eqnarray}
with the initial conditions
\begin{equation}
  \label{wp_3}
  a_0(\rho,0)=\Theta(-\rho)\;,\qquad b_0(\rho,0)=\Theta(\rho)\;,
\end{equation}
where $\Theta$ is the Heaviside function, 
and the boundary conditions
\begin{equation}
  \label{wb_3}
\left\{
  \begin{array}{l}
    a_0(-\infty,\tau)=1\;,\\
    a_0(+\infty,\tau)=0\;,
  \end{array}
  \right. \qquad
\left\{
  \begin{array}{l}
    b_0(-\infty,\tau)=0\;,\\
    b_0(+\infty,\tau)=1\;.
  \end{array}
  \right.
\end{equation}
For $n=1,2,3,\ldots$ we obtain the equations for the functions of n-th order
\begin{eqnarray}
  \frac{\partial^\alpha a_n(\rho,\tau)}{\partial \tau^{\alpha}}&=&\frac{\partial^2 a_n(\rho,\tau)}{\partial\rho^{2}}-R_n(\rho,\tau)\;,\label{eq9a}\\
  \frac{\partial^\alpha b_n(\rho,\tau)}{\partial \tau^{\alpha}}&=&\frac{\partial^2 b_n(\rho,\tau)}{\partial \rho^{2}}-R_n(\rho,\tau) \;,\label{eq9b}
\end{eqnarray}
where 
\begin{equation}\label{}
  R_n(\rho,\tau)=\sum_{k=0}^{n-1}a_k(\rho,\tau)b_{n-k}(\rho,\tau)\;,
\end{equation}
with the initial condition
\begin{equation}
  \label{wp_1a}
  a_n(\rho,0)\equiv b_n(\rho,\tau)\equiv 0\;,\quad n\geq1\;,
\end{equation}
and the bonduary ones
\begin{equation} 
  \label{wb_1a}
  a_n(-\infty,\tau)=a_n(+\infty,\tau)=0\;,\qquad b_n(-\infty,\tau)=b_n(+\infty,\tau)=0\;.
\end{equation}
We solve Eqs.~(\ref{eq:5a}), (\ref{eq:5b}), (\ref{eq9a}) and~(\ref{eq9b}) by means of the Laplace transform method \cite{k}. The zeroth order equations~(\ref{eq:5a}) and~(\ref{eq:5b}) appear to be subdiffusion equations without chemical reactions. Equations~(\ref{eq9a}) and~(\ref{eq9b}) become even more difficult to solve when the order of the perturbation method increases. We will find the exact solutions for $a_i$ and $b_i$ where $i=0,1$. When the parameter $\epsilon$ is small ($\epsilon\ll 1$), the approximate solutions to Eqs.~(\ref{eq:2a}) and~(\ref{eq:2b})  are assumed to be
\begin{equation}
  \label{rozw}
  a=a_0+\epsilon a_1\;,\qquad b=b_0+\epsilon b_1\;.
\end{equation}

\section{Approximate solution to subdiffusion--reaction equations}

The calulations are arduous, thus the details of these calculations will be presented elsewhere.
In term of dimensional variables the solutions are
\begin{equation}\label{eq:61}
A^\pm(x,t)=A_0^\pm(x,t)+A_1^\pm(x,t)\;,
\end{equation}
\begin{equation}\label{eq:61a}
B^\pm(x,t)=A^\pm(-x,t)\;,
\end{equation}
where
\begin{eqnarray}
A_0^-&(x,t)=&C_0\left[1-\frac{1}{2}\sum_{i=0}^\infty{\frac{1}{\Gamma(1-\alpha k/2)k!}\left(\frac{x}{\sqrt{D}t^{\alpha/2}}\right)^i}\right]\;,\label{eq:62}\\
A_0^+&(x,t)=&\frac{C_0}{2}\sum_{i=0}^\infty{\frac{1}{\Gamma(1-\alpha k/2)k!}\left(-\frac{x}{\sqrt{D}t^{\alpha/2}}\right)^i}\;.\label{eq:63}
\end{eqnarray}
The function $A_1^\pm(x,t)$ is controlled both by chemical reactions and diffusion. It can be treated as a corrections of $A_0^\pm(x,t)$ which is controlled by the diffusion process only. Thus, $A_1^\pm$ makes samller the particles concentration described by the pure diffusion solution and has a proper physical meaning for negative values only, and it is expressed by the formula
\begin{equation}\label{eq:64}
A_1^\pm(x,t)= C_0^2kt^\alpha F^\pm(x,t)\;,
\end{equation}
when $F^\pm(x,t)\leq 0$, but we put $A_1^\pm(x,t)=0$ when $F^\pm(x,t)> 0$, where
\begin{equation}\label{eq:65}
F^\pm(x,t)-\frac{1}{4}\sum_{i=0}^\infty{Q_i\left(\frac{x^2}{Dt^\alpha}\right)^i}+\sum_{i=1}^\infty{P_i^\pm \left(\frac{x}{\sqrt{D}t^{\alpha/2}}\right)^i}\;,
\end{equation}
\begin{eqnarray}
Q_i&=&\frac{1}{(2i)!\Gamma(1-(i-1)\alpha)}\;,\;\label{eq:66}\\
P_{i}^\pm &=&\frac{1}{i!\Gamma(1-\alpha(i/2-1))}\sum_{m=0}^{\left\lfloor (i+1)/2\right\rfloor-1}{m!e^\pm_{m}}\;,\;\label{eq:67}
\end{eqnarray}
$\left\lfloor u\right\rfloor$ denotes here an integer part of a number $u$ ($\left\lfloor u\right\rfloor\leq u$),
$e^\pm_0=1/4$, $e^\pm_1=0$, and for $m\geq 2$ we get
\begin{eqnarray}
e^-_m&=&\frac{1}{4k!}\left[1-\sum_{j=1}^m{\frac{\Gamma(1-\alpha m/2)m!}{\Gamma(1-\alpha j/2)j!\Gamma(1-\alpha(m-j)/2)(m-j)!}}\right]\;,\label{eq:48}\\
e^+_m&=&\frac{(-1)^m}{2m!}\left[1-\frac{1}{2}\sum_{j=1}^m{\frac{\Gamma(1-\alpha m/2)m!}{\Gamma(1-\alpha j/2)j!\Gamma(1-\alpha(m-j)/2)(m-j)!}}\right]\;\label{eq:49}.
\end{eqnarray}

\begin{figure}[h!]
  \begin{center}
    \includegraphics[scale=0.4]{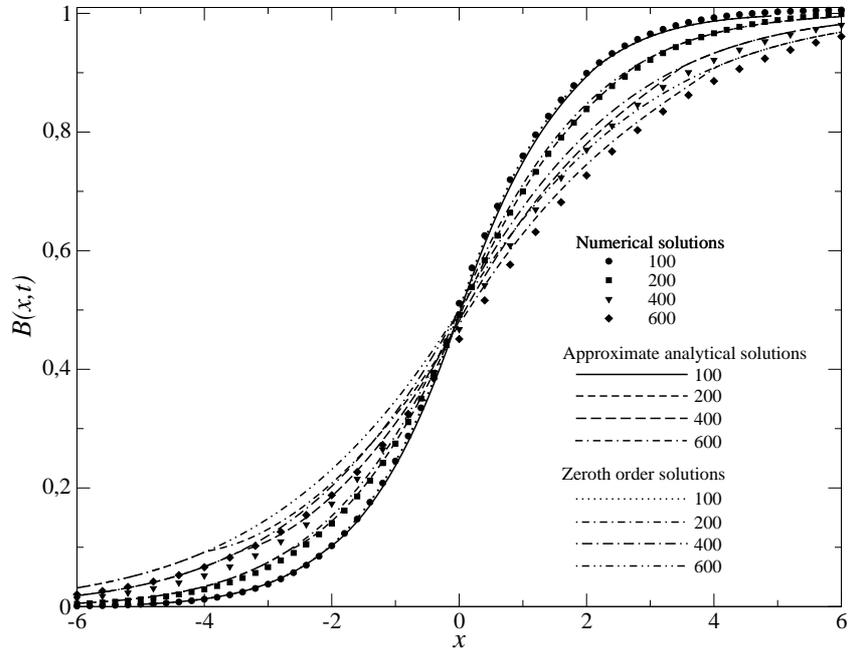}
    \caption{\label{Fig.1} Concentration profiles calculated for $\alpha=0.7$, $D=0.05$, $C_0=1$, $k=0.001$ and for the times given in the legend; symbols represent the numerical solutions, lines --- the approximate analytical solutions (\ref{eq:61a}) and the zeroth order solutions (i.e. for the system without chemical reactions).}
  \end{center}
\end{figure}

\begin{figure}[h!]
  \begin{center}
    \includegraphics[scale=0.4]{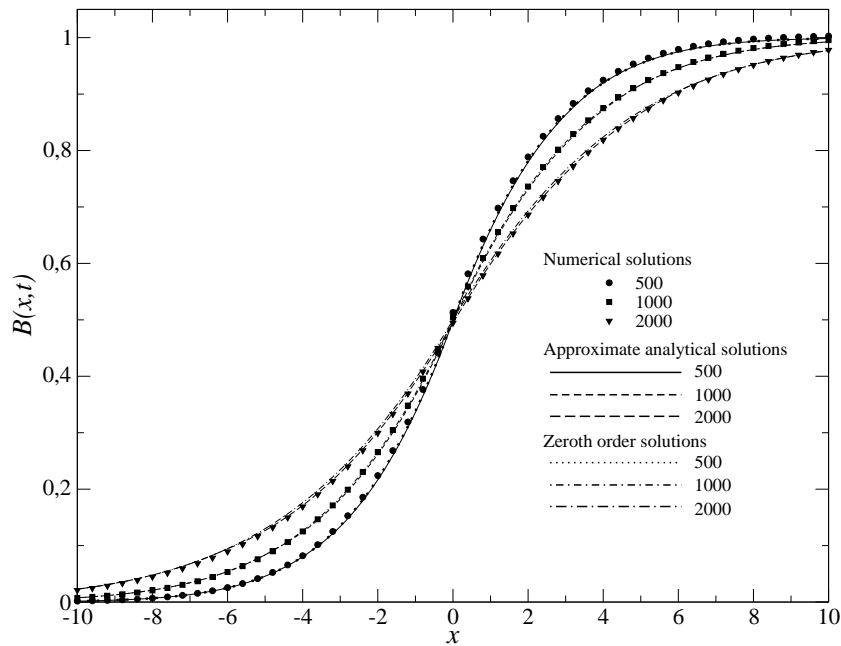}
    \caption{\label{Fig.2} The same situation as in Fig.~\ref{Fig.1} but for $\alpha=0.7$, $D=0.05$, $C_0=1$, $k=0.0001$.}
  \end{center}
\end{figure}

Let us note, that in the functions~(\ref{eq:61})--(\ref{eq:49}) the parameter $\epsilon$ is absent. This absence is caused by making the reciprocal inverse transformation of the variables (from $(x, t)$ to $(\rho, \tau)$ and vice versa). As we have mentioned above, we expect that the perturbation method gives the accurate solutions for dimensionless equations~(\ref{eq9a}) and~(\ref{eq9b}) when $\epsilon\ll 1$. However, it is not clear whether after transforming the dimensionless solutions into the dimensional ones these solutions will be correct. In order to ensure this, we compare the solutions obtained by the perturbation method~(\ref{eq:61a}) with the numerical solutions to the subdiffusion--reaction equations~(\ref{eq3:a}) and~(\ref{eq3:b}). The numerical procedure of solvnig the subdiffusion equations is presented in~\cite{kl1,kl2}. In Figs.~\ref{Fig.1} and~\ref{Fig.2} we present the comparision between the analytical and numerical solutions. We observe that this agreement is reasonably accurate for the times presented in the figures. However, as we can see in Figs.~\ref{Fig.1} and~\ref{Fig.2} the similarity between the approximate solutions and the numerical ones decreases with ascending time. The zeroth order solutions corresponding to the solution to pure subdiffusion equations do not match the numerical solutions accurately but after adding the correction of the first order this match is considerably improved.

\section{Final remarks}

In this paper we find the approximate solution to subdiffusion--reaction equations~(\ref{eq3:a}) and~(\ref{eq3:b}) by means of the perturbation method. We calculate this solution for the zeroth and the first order of the perturbation method alone. We also compare our analytical solutions with the numerical ones. This comparison is presented in Figs.~\ref{Fig.1} and~\ref{Fig.2}. The considerations presented in our paper show that the perturbation method can be useful in solving nonlinear subdiffusion--reaction differential equations with fractional time derivatives.

\section*{Acknowledgments}
This paper was partially supported by the Polish National Science Centre under grant No. 1956/B/H03/2011/40.

\end{document}